\documentclass[]{pasj01}
\usepackage{ulem}
\begin{document} 
\Received{}%{yyyy/mm/dd}
\Accepted{}%{yyyy/mm/dd}
%\Published{yyyy/mm/dd}

\title{Repeating and Non-repeating Fast Radio Bursts from Binary Neutron Star Mergers}

%%% begin:list of authors
% Do NOT capitalize all letters in "textsc".
\author{Shotaro \textsc{Yamasaki}\altaffilmark{1}%
}
\altaffiltext{1}{Department of Astronomy, School of Science, The University of Tokyo,  7-3-1 Hongo, Bunkyo-ku, Tokyo 113-0033, Japan}
\email{yamasaki@astron.s.u-tokyo.ac.jp}

\author{Tomonori \textsc{Totani},\altaffilmark{1,2}}
\altaffiltext{2}{Research
Center for the Early Universe, School of Science, The University of
Tokyo, 7-3-1 Hongo, Bunkyo-ku, Tokyo 113-0033, Japan}
%\email{bbbbb@xxx.xxx.xx.xx}

\author{Kenta \textsc{Kiuchi}\altaffilmark{3}}
\altaffiltext{3}{Center for Gravitational Physics, Yukawa Institute for Theoretical Physics, Kyoto University, Kyoto 606-8502, Japan}
%\email{ccccc@xxx.xxx.xx.xx}
%%% end:list of authors

%% `\KeyWords{}' always has to be placed before `\maketitle'.
 %Do NOT move this preamble from here!

\maketitle

\begin{abstract}
Most of fast radio bursts (FRB) do not show evidence for repetition, and
such non-repeating FRBs may be produced at the time of a merger of
binary neutron stars (BNS), provided that the BNS merger rate is close
to the high end of the currently possible range.  However, the merger
environment is polluted by dynamical ejecta, which may prohibit
the radio signal to propagate. We examine this by using a
general-relativistic simulation of a BNS merger, and show that the
ejecta appears about 1 ms after the rotation speed of the merged
star becomes the maximum. Therefore there is a time window in which an
FRB signal can reach outside, and the short duration of non-repeating
FRBs can be explained by screening after ejecta formation. A fraction
of BNS mergers may leave a rapidly rotating and stable neutron star, and
such objects may be the origin of repeating FRBs like FRB 121102.  We
show that a merger remnant would appear as a repeating FRB in a time
scale of $\sim$1--10 yrs, and expected properties are consistent with
the observations of FRB 121102.  We construct an FRB rate evolution
model including these two populations of repeating and non-repeating
FRBs from BNS mergers, and show that the detection rate of repeating
FRBs relative to non-repeating ones rapidly increases with improving
search sensitivity. This may explain that the only repeating FRB
121102 was discovered by the most sensitive FRB search with Arecibo.
Several predictions are made, including appearance of a repeating FRB
1--10 years after a BNS merger that is localized by gravitational wave
and subsequent electromagnetic radiation.
\end{abstract}

\section{Introduction}
\label{sec:intro}

The enigmatic millisecond-duration radio transients, the so-called
fast radio bursts (FRBs) were first discovered by \citet{Lorimer2007},
then confirmed with additional four bursts by \citet{Thornton2013},
and now it is an intensive field of research in astronomy.  About 20
FRBs have been reported to date \citep{Petroff2016}, but their origin
and physical mechanism still remain mysterious.  Their dispersion
measures (DMs) ${\rm DM}\equiv \int n_e dl=300$--$1500{\;\rm
  pc\;cm^{-3}}$ \citep{Petroff2016} are much larger than those
expected for objects in the Milky Way, and a cosmological distance
scale of $z \sim$ 1 is inferred if the dominant contribution to DMs is
from electrons in ionized intergalactic medium (IGM). Counterparts in other
wavelengths (e.g., \cite{Yamasaki2016}) or host galaxies have not yet been detected in most cases.
\citet{Keane2016} reported a radio afterglow of FRB 150418 and
identification of an elliptical host galaxy at $z = 0.492$, but there
is a claim that the radio afterglow may be an AGN activity that is not
related to the FRB \citep{Williams&Berger2016}. Further radio monitoring with high resolution will be needed to settle these disputes \citep{Bassa2016}.

The majority of FRBs do not show evidence for repetition, in spite of
the fact that some of them have been intensively monitored to search
possible repeating bursts \citep{Lorimer2007,Petroff2015}.  The only
exception is FRB 121102, which is the only FRB discovered by the
Arecibo observatory \citep{Spitler2014} and later found to repeat
\citep{Spitler:2016uq,Scholz:2016rpt}. The repetition allowed
sub-arcsecond localization and the first unambiguous identification of
the host galaxy (\cite{Chatterjee2017,Marcote2017,Tendulkar2017}). FRB
121102 was discovered by a high-sensitivity search of Arecibo, and its
burst flux ($\sim$0.02--0.3 Jy) is smaller than that of other FRBs
($\sim$0.2--2 Jy) mostly detected by the Parkes observatory
\citep{Spitler:2016uq}.  If we take into account distance ($z = 0.193$
for FRB 121102 and the DM-inferred redshifts of $z=0.5$--$1.0$ for
other FRBs), the absolute luminosity of FRB 121102 is two orders of
magnitude smaller than other FRBs.  This implies a possibility that
FRB 121102 belongs to a different population from other FRBs.

There is a variety of progenitor models proposed for FRBs; some of
them are related to repeatable populations, while others to
catastrophic events. The former includes giant flares from soft
gamma-ray repeaters (SGRs;
\cite{Popov2010,Thornton2013,Lyubarsky2014,Kulkarni2014}), giant radio
pulses from pulsars \citep{Connor2016,Cordes&Wasserman2016}, repeating
FRBs from a young neutron star
(\cite{Kashiyama&Murase2017,Metzger2017,Beloborodov2017}), collisions
of asteroids with a neutron star \citep{Geng&Huang2015,Dai2016}, and
pulsars interacting with plasma stream \citep{Zhang2017}. The latter
includes binary neutron star (or black hole) mergers
\citep{Totani2013,Mingarelli2015}, binary white dwarf mergers
\citep{Kashiyama2013}, binary black hole mergers \citep{Liu2016} and collapsing supermassive neutron stars
\citep{Falcke&Rezzolla2014}.

In this paper we consider mergers of binary neutron stars (BNS, i.e.,
a binary of two neutron stars) as a possible source of
FRBs. Apparently non-repeating FRBs can be explained by radio emission
at the time of merger.  The exceptionally intense FRB 150807 shows a
small amount of rotation measure (RM) implying negligible
magnetization in the circum-burst plasma \citep{Ravi2016}, which may
favor the clean environment expected around BNS mergers. There is
still a large uncertainties in both FRB and BNS merger rates, but the
FRB event rate is close to the high end of the plausible range of BNS
merger rate, $1 \times 10^4 {\rm \ Gpc^{-3} \ yr^{-1}}$
\citep{Abadie2010}. The latest upper bound on the BNS merger rate by
LIGO \citep{Abbot2016} is also close to this: $1.26 \times 10^4 {\rm
  \ Gpc^{-3} \ yr^{-1}}$ (90\% C.L.), indicating that a BNS merger
should be detected soon if non-repeating FRBs are produced by BNS
mergers \footnote{Shortly after the submission of
    this work, the first gravitational wave event GW170817 from a
    binary neutron star merger was reported, and the BNS merger
rate is estimated to be 
$1540_{-1220}^{+3200} {\rm \ Gpc^{-3} \ yr^{-1}}$
\citep{Abbot2017a}.}. The observed FRB flux can be
explained by magnetic braking luminosity and a radio conversion
efficiency similar to pulsars \citep{Totani2013}.  \citet{Wang2016}
investigated radio emission based on the unipolar inductor model
(\cite{Piro2012,Lai2012}, see also \cite{Hansen2001}).

A theoretical concern of the BNS merger scenario is, however, that the
environment around the merger would be polluted by
matter dynamically expelled during the merger process, which may
prohibit the radio signal to be transmitted.  The first aim of this
work is to investigate this issue by using a numerical-relativity
simulation of a BNS merger. We will compare the rise of rotation power
that may produce an FRB and the timing of dynamical matter ejection, and
examine whether there is a time window in which an FRB is produced and
transmitted to an observer.

It is obvious that a radio burst at the time of a BNS merger cannot
explain the repeating FRB 121102. A young neutron star possibly with
strong magnetic field (i.e., a magnetar) is then popularly discussed
as the source of FRB 121102, which is surrounded by a pulsar wind
nebula (PWN) that is responsible for the observed persistent radio
emission. Therefore a core-collapse supernova, especially in the class
of superluminous supernovae (SLSNe), is discussed as the progenitor of
FRB 121102, because formation of a rapidly rotating magnetar is one of
the possible explanations for the extreme SLSN luminosity, and because
of the host galaxy properties (dwarf and low metallicity)
(\cite{Kashiyama&Murase2017,Metzger2017}).

However, a fraction of BNS mergers may leave a massive neutron star
which is indefinitely stable or temporarily stable by a rotational
support (e.g., \cite{Gao2013,Metzger&Piro2014,Piro2017}).  The
fraction strongly depends on the equation of state (EOS) for nuclear
matter as well as neutron star mass distribution, 
which may be either negligible or the
majority \footnote{
After the detection of GW 170817, there have been
several attemps to constrain the nature of the
merger remnant and the maximum mass of neutron stars,
but an unambiguous conlcusion has not yet been obtained
and the fraction of merger events leaving a long-lived
neutron star is still highly uncertain. }.  The
latter requires that EOS is stiff enough to support a spherical
neutron star with the maximum mass of $\gtrsim 2.7 M_\odot$.  Such
remnant neutron stars should be rapidly spinning by the large angular
momentum of the original binary, and their magnetic field can be
amplified by the merger process (e.g., \cite{Kiuchi2014}), possibly to
the magnetar level. The ejecta mass from a BNS merger is much smaller
than SLSNe, making the transmission of radio signal easier.  The
estimated event rate of BNS mergers is higher than that of SLSNe by
1--2 orders of magnitude (\cite{Abadie2010,Quimby2013}), and hence the
production rate of rapidly rotating neutron stars by BNS mergers may
be higher than that by SLSNe.

The second aim of this work is to examine merger-remnant neutron stars
as the origin of repeating FRBs like FRB 121102.  We make
order-of-magnitude estimates of various physical quantities and
compare with the observational constraints for FRB 121102. We then
propose a unified scenario for repeating and non-repeating FRBs from
BNS mergers. Non-repeating and bright FRBs are produced as a single
catastrophic event at the time of merger, while repeating and faint
FRBs are produced by young and rapidly rotating neutron stars left after BNS
mergers.  We then present an FRB rate evolution model including these
two populations, and examine the relative detection rate as a
function of search sensitivity. This may give a hint to explain the
fact that the only repeating FRB was detected as the faintest FRB.

The outline of this paper is as follows.  In section \ref{sec:
  Simulation & Method}, we describe the details of the BNS simulation
used in this work.  We then examine ejecta formation by the merger and
discuss the possibility of producing a non-repeating FRB in section
\ref{sec: Result}. The merger-remnant neutron star scenario for
repeating FRBs is compared with the available observational
constraints in section \ref{sec:repeating-FRBs}, and the FRB rate
evolution model is presented in section \ref{sec:rate}.
Conclusions will be given with some discussions in section \ref{sec:
  Conclusions}. The adopted cosmological parameters for a flat universe
are $H_0 = 67.8\,{\rm km\, s^{-1}\,Mpc^{-1}}$, 
$\Omega_{\rm M} = 0.308$, and 
$\Omega_{\Lambda}= 0.692$ \citep{Planck2016}.

\section{BNS Merger Simulation}
\label{sec: Simulation & Method}

Methods of the BNS merger simulation used in this work are presented
in \citet{Kiuchi2014}. The simulation employs the moving puncture
gauge, and the spatial coordinates (denoted by $xyz$) are defined so
that they asymptotically become the Cartesian coordinate system
towards a point at infinity from the center.  The simulation is
performed in a cubic box and the centers of two neutron stars are
located in the $z=0$ plane.  A reflection symmetry with respect to the
$z=0$ plane is assumed.  A fixed mesh-refinement algorithm with seven
levels is adopted for the spatial coordinates to resolve the wide
dynamic range of a BNS merger. The outermost (i.e., the first level)
box has $469 \times 469 \times 235$ grids in $x$-$y$-$z$, with a grid
size of $\sim$ 9.6 km. (The number of $z$-direction grids is only for
the upper half of the cube.)  In the second level, the box of half
size (i.e., 1/8 in volume) with the same box center is simulated with
a two times finer grid size, while the grid number is the same. This
is repeated in the same way to the seventh level where the mesh size
is $1/2^6$ of the first level ($\sim$ 150 m).

We employ the H4 EOS of \citet{PhysRevLett.67.2414}, with which the
maximum mass of neutron stars is $2.03 \:M_{\odot}$. Two neutron stars
have the same ADM mass of 1.35 $M_{\odot}$ when they are
isolated. This simulation does not include magnetic fields; there are
no known BNS simulations in which dynamical ejecta mass is
significantly changed by the effects of magnetic fields.  The
simulation starts with an orbital angular velocity $\Omega_{\rm orb}
\sim 1.7 \times 10^3$ s$^{-1}$ and a binary separation of $\sim$ 50
km, and the merger occurs after several in-spiral orbits.  The
simulation finishes at 15 ms after the merger, and at that time the
merged hypermassive neutron star (HMNS) is still rotationally
supported against a gravitational collapse.

Our purpose is to investigate the time evolutionary properties of matter
ejected to outer regions, and we do not have to examine quantitatively the
general relativistic effects that are important around the merger
center.  Therefore in this work we present physical quantities
assuming that the simulation grids are on the classical Cartesian
coordinate system throughout the box, and the simulation time grids
are on the classical time coordinate.

For computational reasons, numerical simulations of a BNS merger
usually set an artificial atmosphere around stars, and in our
simulation the density of atmosphere is $10^3\:\rm{g\,cm^{-3}}$ within
$r \le 70$ km and it decreases as $\propto r^{-3}$ in outer regions,
where $r$ is the radius from the simulation center. The central
density of the atmosphere is $10^{12}$ times smaller than the nuclear
matter density found inside the neutron stars. Furthermore, the
minimum density that the simulation can reliably resolve is $\sim
10^8\:\rm{g\,cm^{-3}}$. Therefore we consider only matter whose
density is higher than the threshold value, $\rho_{\rm th}=
10^8\:\rm{g\,cm^{-3}}$, when calculating the rest-mass column density
of ejecta.  Even if calculated column density allows transmission of
FRB signals, we cannot exclude a possibility that lower density
material than $\rho_{\rm th}$ absorbs FRB signals.  However, our
results shown below indicate that the matter density rapidly drops at
a certain radius from the HMNS, and material of $\rho < \rho_{\rm th}$
would unlikely affect our main conclusions.

\section{Environment around the merger and FRB possibility}
\label{sec: Result}

\begin{figure*}
 \includegraphics[scale=0.45]{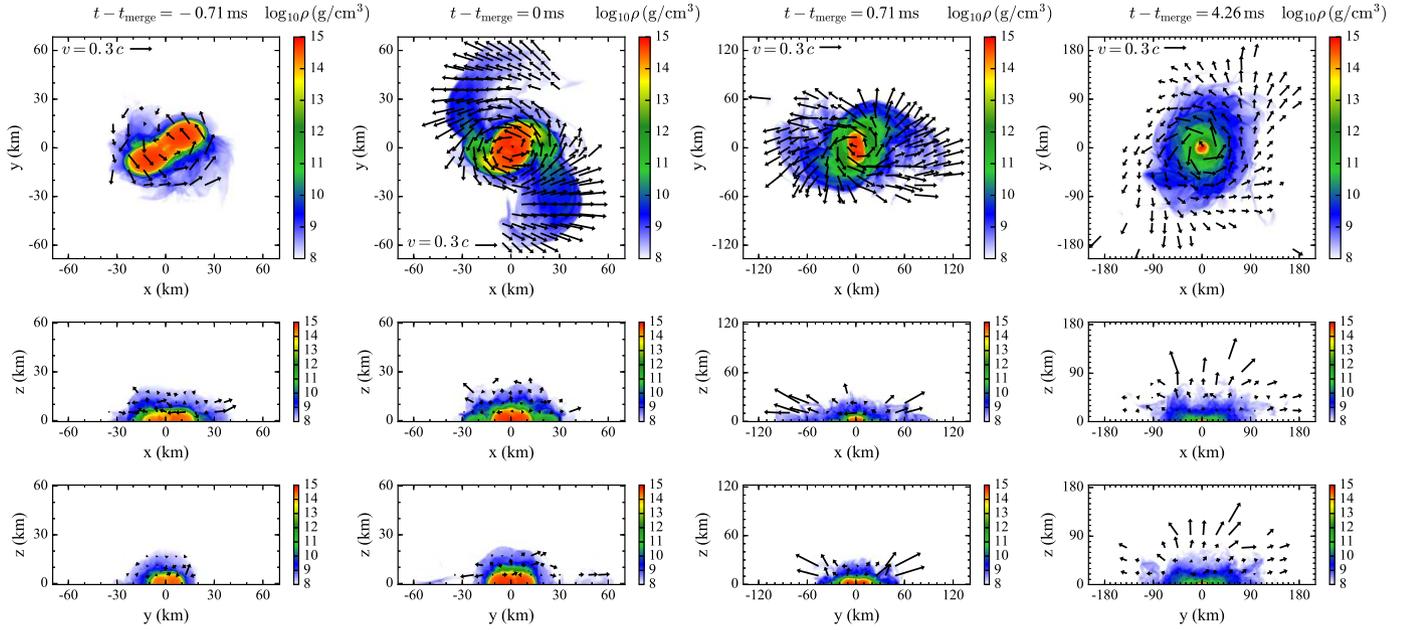}
 \caption{Time snapshots of density contours for the binary neutron
   star merger simulation used in this work, in the $xyz$ coordinates
   which are approximately the classical Cartesian coordinates.  The
   velocity fields are also shown by black vectors.
}
 \label{fig: snapshots}
\end{figure*}

\begin{figure*}
 \includegraphics[scale=0.6]{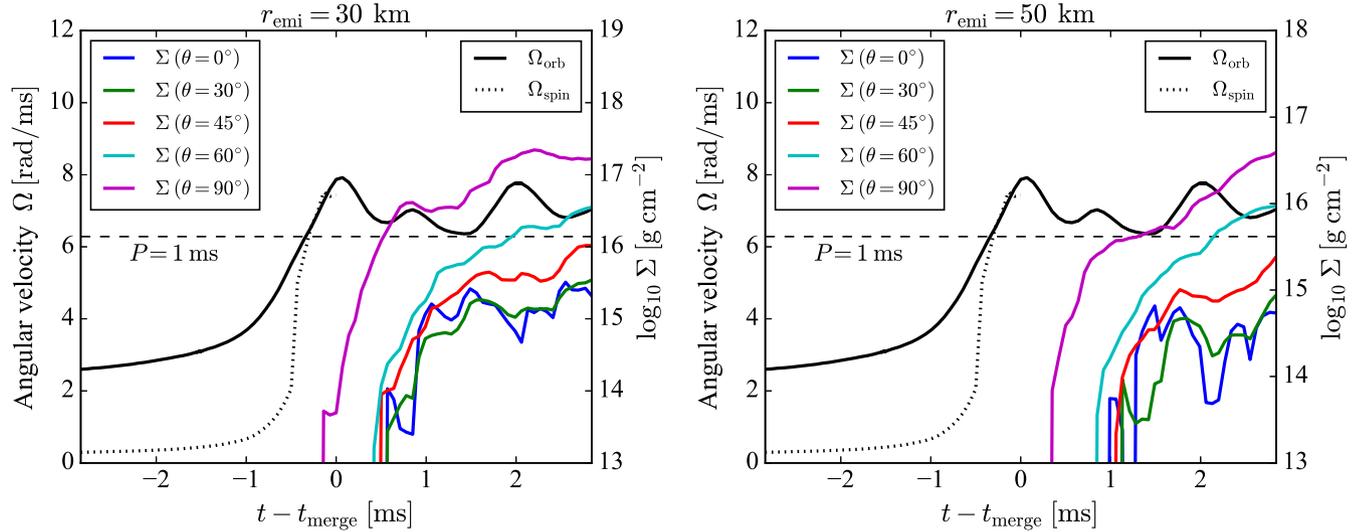}
 \caption{Evolution of angular velocities of orbital motion
   ($\Omega_{\rm orb}$) and spin of each neutron star ($\Omega_{\rm
     spin}$) are shown by solid and dotted black curves, respectively
   (see the left hand ordinate for labels).  The horizontal dashed
   line indicates $\Omega$ corresponding to a rotation period of $1$
   ms. Color curves show rest-mass column density $\Sigma$ in regions
   of $r > r_{\rm emi}$ towards the direction polar angle $\theta$
   from the $z$ axis, for several values of $\theta$. The left and
   right panels are for $r_{\rm emi}$ = 30 and 50 km, respectively,
   and a median about the azimuth angle $\phi$ is taken for
   $\Sigma$. }
 \label{fig: evolution of spin rate & column density}
\end{figure*}

\subsection{Orbital evolution and neutron star spin-up}

Figure \ref{fig: snapshots} presents time snapshots of density
contours and velocity fields of the simulation, spanning from 0.71
ms before to 4.26 ms after the merger, where the merger time $t_{\rm
  merge}$ is defined as the time when the density peaks of the two
neutron stars merge into one.

FRBs are expected to be generated by rotation, either the orbital
motion of the two neutron stars or spins of individual neutron
stars. The rotation angular velocity of the orbital motion
($\Omega_{\rm orb}$) and that of the individual neutron star spin
($\Omega_{\rm spin}$) are shown as a function of time in figure \ref{fig: evolution of spin rate & column density}. For this
calculation, we first calculate the angular velocity at each grid
from the rotation-direction component of fluid velocity, as
\begin{eqnarray}
\Omega({ \bf r}) = \frac{1}{ | {\bf r}_{xy} | } \, 
{\bf v}  \cdot \left( \frac{ {\bf n}_z \times
{\bf r} }{ | {\bf n}_z \times {\bf r} | }
\right) \ ,
\end{eqnarray}
where ${\bf r}$ is the position vector measured from the rotation
center on the $z=0$ plane, ${\bf v}$ the three velocity of fluid,
${\bf n}_z$ the unit spatial vector to the $z$ direction, and ${\bf
  r}_{xy}$ the projection of ${\bf r}$ onto the $xy$ plane. Then the
average rotation velocity is calculated as the mass-weighted mean:
\begin{eqnarray}
\Omega_{\rm av} = \frac{\int \Omega({\bf r}) \rho({\bf r}) d{\bf r}  }{
\int \rho({\bf r}) d{\bf r} } \ ,
\end{eqnarray}
where $\rho$ is the rest-mass density.  The orbital rotation velocity
$\Omega_{\rm orb}$ is simply calculated by setting the rotation center
at the center of the simulation box and integration over the whole
simulation box. The spin of each neutron star $\Omega_{\rm spin}$ is
calculated by setting the rotation center at the density peak of one
of the two neutron stars. For the integration region of $\Omega_{\rm
  spin}$, we separate the simulation box into two by a plane including
the simulation box center and perpendicular to the line connecting the
two centers of the neutron stars. Then the integration of $\Omega_{\rm
  spin}$ is performed only over the half side including the neutron
star considered.  As expected, $\Omega_{\rm spin}$ of the two neutron
stars are almost the same, and it becomes the same as $\Omega_{\rm
  orb}$ after the merger.

Though the calculation of $\Omega_{\rm orb}$ is completely Newtonian,
we compare this with the angular frequency of the dominant quadrupole
mode of gravitational wave radiation ($\Omega_{\rm gw}$) calculated
from the simulation by a relativistic method using the Weyl scalar
\citep{Yamamoto2008}.  We then confirmed that the expected relation,
$\Omega_{\rm orb} = \Omega_{\rm gw}/2$, holds within a 10\% accuracy.

It can be seen in figure \ref{fig: evolution of spin rate & column
  density} that $\Omega_{\rm orb}$ gradually increases to the merger,
but $\Omega_{\rm spin}$ suddenly rises up at $\sim$ 0.5 ms before the
merger. Our simulation does not include viscosity, and hence a tidal
lock by viscosity does not occur. A tidal lock is not expected to
occur even if viscosity is taken into account \citep{Bildsten1992}.
After the sharp rise, $\Omega_{\rm spin}$ is almost the same as
$\Omega_{\rm orb}$, and energy for FRBs can be extracted by the spin
of magnetic fields of the merging star with a rotation period of about
1 ms. A coherent dipole magnetic field may be that of neutron stars
before the merger, or may be formed during the merger process.  The
energy loss rate of the dipole emission formula is proportional to
$\Omega_{\rm spin}^4$, and hence the chance of producing an FRB
rapidly increases at $\sim$ 0.5 ms before the merger.

\subsection{Ejecta formation}

Next we consider ejecta distribution. Figure \ref{fig: evolution of spin rate & column density} shows the time evolution of the
rest-mass column density,
\begin{eqnarray}
\Sigma(\theta, \phi \, ; \, r_{\rm emi})
= \int_{r_{\rm emi}}^\infty \rho({\bf r}) dr \ , 
\end{eqnarray}
which is integrated over the radial direction from the simulation
center excluding the inner region of $r < r_{\rm emi}$, where $r$,
$\theta$, $\phi$ are spherical coordinates.  As mentioned in section
\ref{sec: Simulation & Method}, low density grids with $\rho <
\rho_{\rm th}$ are excluded from this calculation.  We show the cases
of $r_{\rm emi}$ = 30 and 50 km, for several values of polar angle
$\theta$ from the $z$ direction. The light cylinder radius becomes
$\sim$50 km for a rotation period of 1 ms, and hence it is reasonable
to expect that FRB radiation occurs at 30--50 km from the center.  The
column density also depends on the azimuth angle $\phi$, and here we
take the median of $\Sigma(\phi_i)$ to show a typical column density,
where $\phi_i$ is the 360 grids in $\phi$ = 0--$2\pi$ to calculate
$\Sigma$.  (We avoid a simple mean because it is biased when a high
density ejecta exist into one direction, though its covering fraction
on the sky is small.)

This figure shows that $\Sigma$ significantly increases 0--1 ms after
the merger. Ejecta to the equatorial directions ($\theta \sim
90^\circ$) appear earlier, because dynamical mass ejection is driven
first by tidal force, and then shock heated components are ejected to
the polar direction from the HMNS \citep{Sekiguchi2015}. Since the
minimum density resolved in the simulation is $10^8 \ \rm g
\ cm^{-3}$, column density of $\Sigma \lesssim 10^{14} \ \rm g
\ cm^{-2}$ cannot be resolved on the scale of 30--50 km.  Once
$\Sigma$ becomes larger than this, there is no chance for an FRB
emission to escape, because the optical depth of electron scattering
is many orders of magnitudes larger than unity.  The rapid increase of
$\Sigma$ by many orders of magnitude occurs at about 1 ms after the
merger to most of directions, implying that the environment
before this is similar to that of isolated neutron stars.

\begin{figure*}
 \includegraphics[scale =0.6]{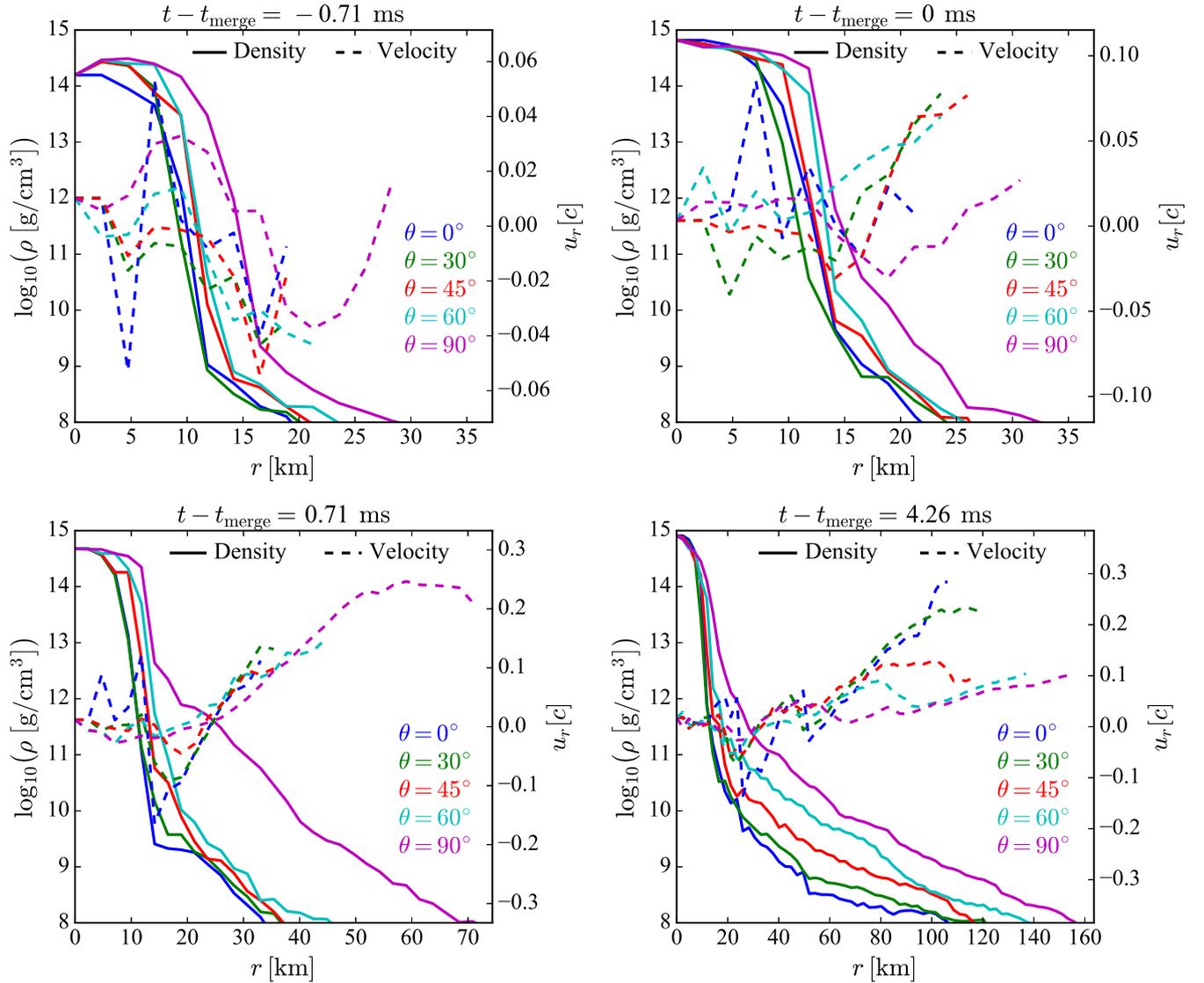}
 \caption{Time snapshots (corresponding to figure \ref{fig: snapshots})
 of radial profiles of rest-mass density
   (solid colored lines) and radial fluid velocity (dashed colored
   lines), for several values of polar angle $\theta$ from the $z$
   axis. These quantities are the median about the azimuthal angle
   $\phi$.  The radius is measured from the merger center.  
}
 \label{fig: radial profiles}
\end{figure*}

Figure \ref{fig: radial profiles} shows time snapshots of radial
profiles of rest-mass density and velocity. Here, again the median is
plotted about the azimuth angle.  Except for the equatorial ($\theta =
90^\circ$) direction, the density sharply drops from $10^{14}$ to
$\sim 10^9 \ \rm g \ cm^{-3}$ at the surface of newly born HMNS. An
extended tail of the density profile at $\rho \sim 10^8 \ \rm g
\ cm^{-3}$ is seen, but it may be an artifact because this low density
is close to the simulation resolution.  Well-resolved ejecta with
$\rho \gg 10^8 \ \rm g \ cm^{-3}$ and positive radial velocity are
seen only into the equatorial direction at the time of merger, and
those into other directions appear a few ms after the merger.  The
ejecta velocity is at most 0.1--0.2$c$, and it takes about 1 ms for
such an ejecta to expand into the outer regions of $r > r_{\rm emi}
\sim$ 30--50 km.

These results imply that a significant ejecta formation and expansion
to the scale of 30--50 km occurs about 1 ms after the merging
neutron stars start to rapidly spin.  Therefore there is a short time
window of $t - t_{\rm merge} \sim -0.5$ to $0.5$ ms in which the
ejecta is not yet formed but the magneto-rotational energy production
rate is sufficiently high to produce an FRB emission. This also gives a
possible explanation for the observed $\sim$1 ms duration of 
non-repeating FRBs.

\section{Repeating FRBs from the
Remnant Neutron Star after a BNS Merger}
\label{sec:repeating-FRBs}

\subsection{Formation of the Remnant Neutron Star and Its Environment}
\label{subsec:PWN}

Hereafter we consider the case that a BNS merger leaves a merged
neutron star that is indefinitely stable without rotation or
rotationally supported for a time scale longer than the repeating FRB
lifetime.  Its initial spin period is $P_i = 2\pi/\Omega_i \sim
1$ ms, mass $\sim 2.6 M_{\odot}$, and radius $R \sim 15$ km.
If there is a loss of rotation energy by
  e.g. gravitational wave emission, initial rotation period may be
  larger. The rotational energy of the star is
\begin{eqnarray}
E_{\rm rot} =\frac{1}{2}I\Omega_{\rm i}^2\approx 9.2 \times10^{52}
\ \ {\rm erg},
\end{eqnarray}
where $I$ is the momentum of inertia of the star.  The spin-down
timescale by magnetic breaking is given as
\begin{eqnarray}
\label{eq: spin-down timescale}
 t_{\rm sd} = \frac{3c^3 I P_i^2}{4\pi^2B_*^2R^6}
\sim 2.7 \; B_{12.5}^{-2} \ \ {\rm yrs}, 
\end{eqnarray}
where $B_{12.5} = B_*/(10^{12.5} \ \rm G)$ is the strength at the star
surface. We adopt $10^{12.5}$ G as a reference value that is typical
for isolated pulsars, but the magnetic field strength may be enhanced
by the merger process to $B\gtrsim10^{13}$ G as suggested by numerical
simulations (e.g., \cite{Kiuchi2014}), and in such a case $t_{\rm sd}$
could be smaller.\footnote{  
 It should be noted that the
  rotation energy $E_{\rm rot}$ is quickly converted into pulsar wind
  if magnetic field is as strong as magnetars ($10^{15}$ G) and the
  neutron star survives longer than the spin-down time.  This is
  excluded for the particular case of GW 170817, because such a large
  energy is not observed. However, a low magnetic field of $\lesssim
  10^{12.5}$ G is not excluded because of the longer spin-down time. }

Consider ejecta mass $M_{\rm ej}
=10^{-2}\,M_{\odot}$ and velocity $\beta_{\rm ej,0}
= \upsilon_{\rm ej, 0} / c = 0.1$, which are within the typical ranges
for a BNS merger. The dynamical ejecta mass decreases with stiffer EOS
(\cite{Hotokezaka2013}), and the ejecta mass may be enhanced by disk
wind (\cite{Shibata2017a}). The ejecta kinetic energy would be changed
when the rotation energy of the newly born massive neutron star is
injected into the merger ejecta in the form of pulsar wind, within a
time scale of $t_{\rm sd}$. Assuming that an energy of $E_{\rm rot}$
is injected as relativistic particles or Poynting flux at a radius
$r_{\rm inj} = \upsilon_{\rm ej, 0} t_{\rm sd}$, we can estimate the
accelerated velocity and internal energy of ejecta (pulsar wind
nebula) to be $\beta_{\rm PWN} = 0.85$ and $U_{\rm
  inj}= 4.0\times10^{52}$ erg, respectively, from
energy and momentum conservation.  This bulk motion
  speed is mildly relativistic, but we ignore the relativistic effect
  for simplicity in the order-of-magnitude analysis.  We assume that
ejecta is freely expanding except for the velocity change at $r =
r_{\rm inj}$. The expanding ejecta would be decelerated by
interstellar medium (ISM) when a comparable ISM has been swept up, but
this effect can be ignored if we consider evolution before the
deceleration time $t_{\rm dec}\sim 15 \, M_{\rm ej,
    -2}^{1/3} \, n_{\rm ISM,\,-3}^{-1/3} \, \beta_{\rm PWN}^{-1} \;
  {\rm yr}$, where $n_{\rm ISM} = 10^{-3} n_{\rm ISM, -3} \ \rm
cm^{-3}$is the ISM density.  In this work we do not consider the interaction with ISM
for simplicity.

 The energy injection by pulsar wind would heat up
  the ejecta matter and also generate magnetic fields in the ejecta,
  $B_{\rm ej}$.  We assume that a fraction $\epsilon_{\rm B}$ of the
  internal energy density $u_{\rm inj}$ is converted into magnetic
  fields at the time of energy injection, as $B_{\rm ej, inj}^2
  /(8\pi)\approx \epsilon_{\rm B}\, u_{\rm inj}$.  After the injection
  $B_{\rm ej}$ evolves by adiabatic expansion and conserved
  magnetic flux, i.e., $B_{\rm ej} = B_{\rm ej, inj}(r/r_{\rm
    inj})^{-2}$.  This reduces to
\begin{eqnarray}
\label{eq:ejecta magnetic field}
B_{\rm ej}= 5.6 \times10^{-2} \ \epsilon_{{\rm B},-2}^{1/2} \;
B_{12.5}^{-1}\;\beta_{\rm PWN}^{-2}\;t_{\rm yr}^{-2}  \ \ {\rm G} \ ,
\end{eqnarray}
where $t_{\rm age} = t_{\rm yr} \rm \ yr$ is the time elapsed from the
merger, and we use $\epsilon_{\rm B,-2} = \epsilon_{\rm B}/10^{-2}$
that is inferred from the magnetization parameter of the Crab nebula
\citep{Kennel&Coroniti1984}. Here we assumed the shell thickness
$\Delta r \sim r$ to calculate $u_{\rm inj}$. The dependence on $B_*$
appears by $t_{\rm sd}$ (smaller $r_{\rm inj}$ for stronger $B_*$).
Note that we made an approximation of $t_{\rm age} \sim
r/\upsilon_{\rm PWN}$, which is exactly valid only when $t_{\rm age}
\gg t_{\rm sd}$.  This does not affect the conclusions in
this section from order-of-magnitude estimates.
This magnetic field strength will be used in the next section
to discuss the energetics of synchrotron radiation and rotation 
measure.

\subsection{Comparison with FRB 121102 observations}

\subsubsection{Free-free absorption}
First we estimate the time scale for the ejecta to become transparent
to the free-free absorption of radio signals.  Before energy injection
by pulsar wind, the opacity becomes less than unity at a time
\begin{eqnarray}
t_{\rm tr}^{\rm ff} &\sim& 
4.2 \left(Z/26\right)f_{\rm ion}^{1/5}\,\nu_{9}^{-2/5}
T_{{\rm e},3}^{-3/10}     \nonumber \\
&& \ \ \times \ M_{{\rm ej},-2}^{2/5} \, (\beta_{\rm ej, 0}/0.1)^{-6/5} \ \rm yr 
\end{eqnarray}
after the merger,
where $\nu_9 \equiv \nu/({\rm GHz})$ is the frequency of the radio
signal, $f_{\rm ion}$ is the ionization fraction, $Z$ the mean atomic
number of matter in ejecta, and $T_{e,3} = T_e/({\rm
    10^3 \ K})$ the temperature of ejecta. Here we
  used $\beta_{\rm ej, 0}=0.1$ as the ejecta velocity before the
  energy injection by the pulsar wind, which is valid when $t_{\rm sd}
  > t_{\rm tr}^{\rm ff}$.  Therefore this $t_{\rm tr}^{\rm ff}$ is a
  conservative upper-limit, and $t_{\rm tr}^{\rm ff}$ can be smaller
  by the accelerated ejecta speed when $t_{\rm sd} < t_{\rm tr}^{\rm
    ff}$.  After the energy injection by pulsar wind, electrons in
the ejecta may have relativistic energies if energy conversion from
ions to electrons is efficient. Since the free-free opacity of relativistic
electrons is reduced compared with non-relativistic ones
\citep{Kumar:2017yiq}, the environment would be transparent after the
energy injection.

\subsubsection{Synchrotron self absorption}

 High energy electrons and positrons produced as the
  pulsar wind would form a nebular after interaction with the ejecta.
  Synchrotron self absorption by these electrons and positrons may
  prohibit early radio signal to transmit \citep{Murase2016,Yang2016}.
  Following \cite{Murase2016}, we assume the injected electron energy
  spectrum of the nebula to be a broken power-law, $dN_{\rm
    inj}/d\gamma_e \propto \gamma^{-q}$ with $q = q_1$ ($<$2) at
  $\gamma_{\rm m}\leq \gamma_e\leq \gamma_{\rm b}$ and $q = q_2$
  ($>$2) at $\gamma_{\rm b}\leq \gamma_e\leq \gamma_{\rm M}$, where
  $\gamma_e$ is the electron Lorentz factor.  This is motivated by
  observations of Galactic pulsar wind nebulae. The break Lorentz
  factor is typically $\gamma_{\rm b}\sim10^4$--$10^6$, and $\gamma_m
  \sim 100 \ll \gamma_b$ and $\gamma_M \gg \gamma_b$.

Equating the synchrotron cooling time and the dynamical timescale
  $t_{\rm dyn}\sim r/v_{\rm PWN}$, the cooling break Lorentz factor is
  found as
\begin{eqnarray}
    \gamma_{c}=3.2\times10^{4}\ \,B_{12.5}^{2}\,\beta_{\rm PWN}^{4}\,t_{\rm yr}^{3}\,\epsilon_{B,-2}^{-1} \ ,
\end{eqnarray}
assuming that the magnetic field strength is given by $B_{\rm ej}$.
Therefore electrons are in the slow cooling regime ($\gamma_m \ll
\gamma_c$) for typical timescales of interest ($\gtrsim$ yr) due to
the large velocity of the ejecta, which is in contrast to the fast
cooling case expected for the SN scenario (e.g.,
\cite{Kashiyama&Murase2017}).  The injection electron spectrum is then
conserved at $\gamma_e < \gamma_c$, and the synchrotron absorption is
dominated by electrons with $\gamma_m < \gamma_e < \gamma_c$.  The
spectrum is normalized so that a fraction $\epsilon_{e} \sim 1$ of the
total internal energy $U$ [$=U_{\rm inj}(r/r_{\rm inj})^{-1}$] is
carried by relativistic electrons and positrons, since it is generally
believed that the pulsar wind is dominated by $e^\pm$
\citep{Kennel&Coroniti1984,Tanaka2013}.  We numerically calculated the
absorption optical depth
\begin{eqnarray}
\label{eq. synchrotron tau}
    \tau_{\nu}^{\rm sa}=\frac{r}{8\pi m_{\rm e}c\, \nu^2}\int_{\gamma_{\rm m}}^{\gamma_{\rm M}}\frac{1}{\gamma_e^2}\frac{dN}{d\gamma_e}\frac{d}{d\gamma_e}\left[\gamma_e^2P_{\rm s}(\nu,\gamma_e)\right]\,d\gamma_e,
\end{eqnarray}
where $P_{\rm s}(\nu,\gamma_e)$ is the synchrotron emitting power and
$dN/d\gamma_e$ is the electron energy spectrum which is the same as
$dN_{\rm inj}/d\gamma_e$ at $\gamma_e < \gamma_c$.  For a parameter
range of $q_1=1$--$1.5$, we find that the nebula becomes transparent
to 1 GHz radio emission at $t_{\rm tr}^{\rm sa} \sim 1$ yr.  

Therefore the environment around a BNS merger would become transparent
for a repeating FRB with a time scale of order years, though the
uncertainty is more than one order of magnitude. After the appearance
of a repeating FRB for an observer, the activity would decrease with
time if the neutron star is already in the spin-down phase.  Then the
highest activity of a repeating FRB would last on a time scale similar
to that of the appearance, i.e., of order years.  It should be noted
that the spin down time would become shorter if $B_*$ is stronger, but
a repeater FRB can be formed even in the case of $t_{\rm sd} < t_{\rm
  tr}$, if the remnant neutron star exists on a time scale longer than
$t_{\rm sd}$ and FRBs are produced using e.g. magnetic field energy.

\subsubsection{The persistent radio source}

The source size of the persistent radio emission from FRB 121102
is limited to 
$\lesssim0.7$ pc \citep{Marcote2017}, and this
gives an upper limit on the age $t_{\rm age}$ of this source. 
Assuming that the ejecta is expanding with $\beta_{\rm PWN}$
from the beginning (i.e., $t_{\rm sd} \ll t_{\rm age}$), 
we find 
\begin{eqnarray}
t_{\rm age} < 2.7  \left( \frac{\beta_{\rm PWN}}{0.85} \right)^{-1}  \ \rm yr, 
\end{eqnarray}
which is comparable with the minimum age $\sim5$ yr of FRB 121102.
Therefore expanding size evolution of the persistent radio source may
be observed in the near future, though a realistic morphology must be
considered for a more quantitative prediction.  The source size may be
smaller if $\beta_{\rm PWN}$ is smaller by a larger ejecta mass, or
$t_{\rm sd}$ is comparable to the age.  Another possibility to make the
size smaller is confinement by dense ISM.

The observed luminosity of the persistent radio emission from FRB
121102 ($1.9 \times 10^{39} \ {\rm erg \ s^{-1}}$ at 10 GHz,
\cite{Chatterjee2017}) can be used to estimate the minimum electron
energy emitting synchrotron radiation. Following the formulation of
\citet{Kashiyama&Murase2017} and the magnetic field strength estimated
above, we find the minimum electron energy as $\sim 1.2 \times 10^{49}
\, t_{\rm yr}^{3} \,\epsilon_{B,-2}^{-1/2} \, \beta_{\rm PWN}^{3/2}
\ {\rm erg}$.  This is sufficiently smaller than the maximum energy
available by the rotation of the merged neutron star, $E_{\rm rot}
\sim 10^{53}$ erg, if the age is less than $\sim 10$ yrs.
Even if there is a significant loss of rotation energy at the
stage of the merger, the rotation energy is still sufficiently
larger if $P_i \lesssim 10$ msec.

\subsubsection{Dispersion and rotation measures}

Next we consider dispersion measure (DM) around the remnant neutron
star. DM of ejecta matter (after energy injection by the pulsar
wind) using the standard formula becomes
\begin{eqnarray}
    {\rm DM}_{\rm ej}\approx 5.2\times10^{-1}\
\, M_{{\rm ej},-2}\,f_{\rm ion}\, \beta_{\rm PWN}^{-2}\,t_{\rm yr}^{-2} 
\  {\rm pc \; cm^{-3}} . 
\end{eqnarray}
The DM
contribution from the host galaxy of FRB 121102 is estimated as ${\rm
  DM}_{\rm host}<55$--$225$ pc ${\rm cm}^{-3}$
\citep{Tendulkar2017,Kokubo2017}, and
the DM changing rate is constrained as $< 2 \;{\rm pc \;cm^{-3}\;
  yr^{-1}}$ \citep{Piro2016}.  These constraints can be
easily met in our model
if the age is larger than $\sim$ 1 yr. 

Though rotation measure (RM) is not yet measured for the repeating FRB
121102,\footnote{After the submission of this work, a
    high ($\sim10^5 \ \rm rad \ m^{-2}$) and variable (10\% decrease
    on a half year) rotation measure of FRB 121102 has been reported
    \citep{Michilli2018}, which is a few orders of magnitude higher
    than our plausible estimate in eq. (\ref{eq: RM_BNS}). However, the high RM may be
    explained if we consider a highly clumpy density structure (e.g.,
    dense nebula filaments), which would enhance magnetic fields. The
    observed short variability timescale may favor a young progenitor
    ($\lesssim10$ yr). Further investigation should be done as future
    work. } it has been observed for some FRBs. \citet{Masui2015}
found a relatively large RM contribution from the host galaxy
($\gtrsim 160 \ \rm rad \ m^{-2}$) of FRB 110523, which favors a dense
and magnetized environment like star forming regions or supernova
remnants. On the other hand, small or negligible RMs from the host
galaxy and IGM were observed for the exceptionally bright FRB 150807
($\lesssim 2 \ \rm rad \ m^{-2}$, \cite{Ravi2016}) and FRB 150215 ($<
25 \ \rm rad \ m^{-2}$, \cite{Petroff2017}), which favor a cleaner
environment.

We can calculate RM of the ejecta matter in our model assuming that
the magnetic field is ordered along the line of sight to an observer,
which becomes
\begin{eqnarray}
\label{eq: RM_BNS}
{\rm RM_{\rm ej}} &\sim& 1.2 \times 10^4 \ 
\ \epsilon_{{\rm B},-2}^{1/2}\,B_{12.5}^{-1} \nonumber  \\ 
&& \times \ M_{{\rm ej},-2}\,f_{\rm
  ion}\, \beta_{\rm PWN}^{-4}\,t_{\rm yr}^{-4} \ \ 
{\rm rad \ m^{-2}}\ \ .
\end{eqnarray}
This can be consistent even with the low RMs of FRB 150807 and FRB
150215 if we take $t_{\rm yr} \sim 10$, though dependence on model
parameters is large. Therefore it is possible that these apparently
non-repeating FRBs are also remnant neutron stars after a BNS
merger, and repeating has not yet been detected because of a search
sensitivity and/or limited monitoring time.  Of course, another
possibility is that these FRBs were produced at the time of a BNS
merger, for which we expect even smaller RM.

It should be noted that here we used the standard classical formulae
for DM and RM calculations. However, electrons may have relativistic
energy after the energy injection from pulsar wind.  The relativistic
effect reduces both DM and RM \citep{Shcherbakov2008}, and hence this
does not affect the consistency between our model and observations.

\subsection{Comparison with the Supernova Scenarios}

Supernovae, especially the class of SLSNe, have been proposed as the
progenitor of a young and rapidly rotating neutron star to produce
repeating FRBs
(\cite{Kashiyama&Murase2017,Metzger2017,Beloborodov2017,Dai2017}). Here
we compare the SN scenario with our BNS merger scenario.  Besides the
event rate difference between SLSNe and BNS mergers mentioned in
section \ref{sec:intro}, a large difference is the ejecta mass of
SLSNe that is much larger than that of BNS mergers. Here we discuss
using typical parameter values of $M_{\rm ej, 1} \equiv M_{\rm ej} /
(10 M_\odot)$ and $\upsilon_{\rm ej, 9} \equiv \upsilon_{\rm ej} /
(10^9 \ \rm cm/s)$ (i.e., an explosion energy of $10^{52}$ erg) for a
SLSN \citep{Metzger2017}.  The difference would be smaller in the case
of ultra-stripped SLSNe ($M_{\rm ej} \sim 0.1 M_\odot$,
\cite{Kashiyama&Murase2017}), 
although the event rate and the ejecta mass are
  highly dependent on how to interpret the light curve of rapidly rising
  transients (e.g., \cite{Drout2014,Arcavi2016}).

Because of the larger mass ejecta and slower speed, it would take a
longer time for the environment to become transparent for radio
signals.  The previous studies about the SLSN scenario then considered a
time scale of 10--100 yrs as the age of FRB 121102. Assuming that the
spin-down time of a newly born neutron star is less than $\sim$ 10
yrs, the rotation energy is decreasing with time and hence we expect
that the BNS scenario has a larger available rotation energy to
produce FRBs than the SLSN scenario. Therefore even if the event rate
of the two populations is the same, we expect brighter and more
active FRBs from neutron stars produced by a BNS merger, and hence a
higher chance of detection.

DM, source size and energetics of persistent radio emission in the
SLSN scenario have been discussed in the previous studies, and they
are consistent with observational constraints of FRB 121102.  Compared
with the BNS merger scenario, DM is larger and hence DM variability
would be stronger, while the persistent radio source size is smaller
and hence the size upper limit is more easily met. RM in the SLSN
scenario has not been discussed in the previous studies, and from our
formulations we find
\begin{eqnarray}
{\rm RM_{\rm ej}} &\sim& 1.7 \times 10^{11}
\ \epsilon_{{\rm B},-2}^{1/2}\,B_{14}^{-1}\, \nonumber \\
&& \times \ M_{{\rm ej},1}\,f_{\rm
    ion}\, v_{{\rm ej},9}^{-4}\,t_{\rm yr}^{-4}
\ \ {\rm rad \ m^{-2}} \ .
\end{eqnarray}
Here we assumed that the ejecta has an internal energy of $\sim
10^{52}$ erg at the time of the energy injection from pulsar wind, but
the velocity is not accelerated because the original supernova kinetic
energy is comparable with the energy injected by the pulsar wind.
This RM is much larger than the maximum RM found for FRB 110523, even
if we assume an age of 100 yrs and a strong stellar magnetic field of
$B_*=10^{14} B_{14}$ G.  This implies that all FRBs cannot be a young
neutron star produced by a SLSN, unless the net magnetization is
largely cancelled by small scale fluctuations of magnetic field
directions.

\section{Rate Evolution of Repeating and Non-repeating FRBs}
\label{sec:rate}

\subsection{Cosmic BNS merger rate evolution}

In order to discuss about the FRB detection rate as a function of a
search sensitivity, we first determine the cosmic BNS merger rate as a
function of redshift.  The comoving volumetric BNS merger rate at a
redshift $z$ [corresponding to a cosmic time $t(z)$] is a convolution
of the comoving star formation rate density $\Psi_{\rm SFR}$ and the
delay time distribution (DTD) of BNS mergers from star formation:
\begin{eqnarray}
{\cal R}_{\rm BNS}(z) =  
\int_{0}^{t(z)}\Psi_{\rm SFR}(t-\tau)f_{\rm D}(\tau)d\tau 
 \ ,
\end{eqnarray}
where $\tau$ is the delay time (the time elapsed from the formation of
a stellar binary to the BNS merger), and $f_{\rm D}(\tau)$ is DTD
normalized per unit mass of star formation. 

\begin{figure}
 \includegraphics[width=\columnwidth]{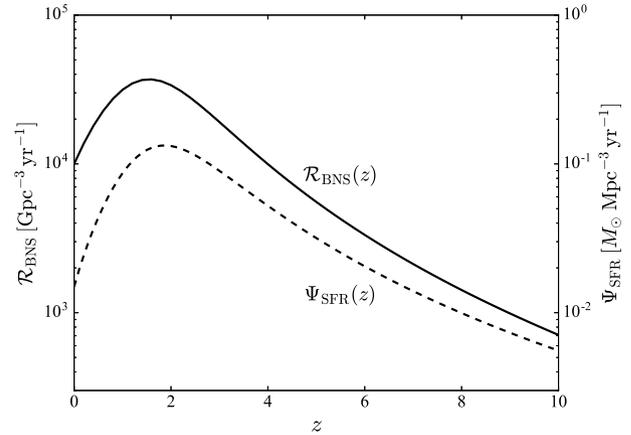}
 \caption{Cosmic star formation rate $\Psi_{\rm SFR}$ (dashed line,
   right-hand-side ordinate) and cosmic BNS merger rate ${\cal R}_{\rm
     BNS}$ (solid line, left-hand-side ordinate) per unit comoving
   volume in our model are shown as a function of redshift $z$.}
 \label{fig: nsns rate}
\end{figure}

We use a functional form of cosmic star formation history,
\begin{eqnarray}
\Psi_{\rm SFR}(z)=0.015 \, 
\frac{ (1+z)^{2.7} }{ 1 + [(1+z)/2.9]^{5.6} } \ M_{\odot}\;{\rm
  yr^{-1}\,Mpc^{-3}}
\end{eqnarray}
in $0<z<8$ derived by \citet{Madau&Dickinson2014}.  DTD of compact
object mergers generally becomes $f_D \propto \tau^{-\alpha}$ with
$\alpha \sim 1$, when it is controlled by gravitational wave radiation
as in the cases of BNS or binary white dwarfs (e.g.,
\cite{Totani2008}).  Here we set $f_D \propto \tau^{-1}$ at $\tau \ge
\tau_{\min}$ and zero otherwise, with $\tau_{\rm min}=10$ Myr, which
is roughly consistent with that calculated by \citet{Belczynski2006}
using a binary population synthesis model.  The BNS merger rate is
normalized as ${\cal R}_{\rm BNS}(0) = 1 \times 10^4 {\rm \ Gpc^{-3}
  \ yr^{-1}}$, which is the ``plausible optimistic'' rate estimate for
local BNS mergers by \citet{Abadie2010}.  It should be noted that the
following results on the ratio of repeating to non-repeating FRB rates
is not affected by this normalization. The calculated 
$\Psi_{\rm SFR}(z)$ and ${\cal R}_{\rm BNS}(z)$ are shown 
in figure \ref{fig: nsns rate}.

\begin{figure}
 \includegraphics[width=\columnwidth]{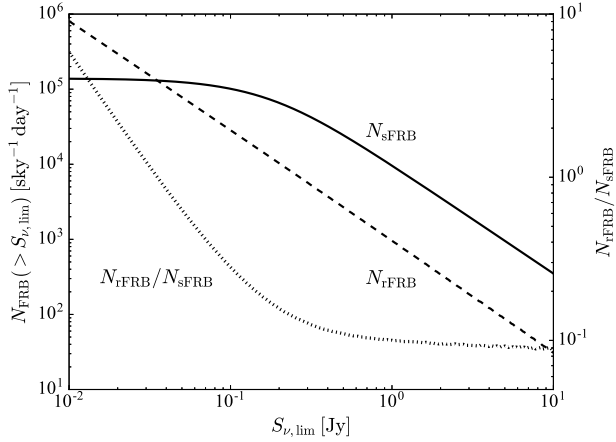}
 \caption{ Occurrence rate of FRBs that are brighter than a search
   flux sensitivity limit, for non-repeating (single) FRBs ($N_{\rm
     sFRB}$) and repeating FRBs ($N_{\rm rFRB}$).  The repeating FRB
   rate is normalized as 10\% of non-repeating FRBs at $S_{\nu, \lim}
   = $ 1 Jy (i.e., $f_r N_r = 400$).  The ratio $N_{\rm rFRB} / N_{\rm
     sFRB}$ is also plotted (see ordinate on the right hand side).}
 \label{fig: logN-logS}
\end{figure}

\subsection{Repeating versus Non-repeating FRB
Detection Rates}

The FRB luminosity function is hardly known, and for simplicity we
adopt the standard candle approximation both for the non-repeating and
repeating populations. There is a large variation in radio spectral
index of FRBs (e.g., \cite{Spitler2014}), and here we simply assume
$L_\nu \propto \nu^0$, and hence $S_\nu \propto (1+z) D_L(z)^{-2}$,
where $L_\nu$ is the absolute FRB luminosity per unit frequency,
$S_\nu$ the observed flux density, and $D_L$ the luminosity
distance. The absolute luminosity is fixed so that $S_\nu = 1.0$ Jy at
$z = 1$ for non-repeating FRBs based on fluxes and DMs observed by
Parkes, while $S_\nu = 0.1$ Jy at $z = 0.19$ for repeating FRBs based
on the case of FRB 121102.

Then the all-sky rates for single (i.e., non-repeating) FRBs ($N_{\rm
  sFRB}$) and repeating FRBs ($N_{\rm rFRB}$) that are brighter than a
limiting flux density $S_{\nu, \lim}$ are calculated using ${\cal
  R}_{\rm BNS}(z)$ as:
\begin{eqnarray}
\label{eq: sFRB rate}
N_{\rm sFRB}(> S_{\nu, \lim}) &=& 
\int_0^{z_{\rm s}} dz \frac{dV}{dz}\frac{{\cal R}_{\rm BNS}}{1+z} \ , \\
\label{eq: rFRB rate}
N_{\rm rFRB}(> S_{\nu, \lim}) &=& 
\int_0^{z_{\rm r}} dz \frac{dV}{dz}\frac{{\cal R}_{\rm BNS}}{1+z}
\, f_r \, N_r \ ,
\end{eqnarray}
where $dV/dz$ is the comoving volume element per unit redshift,
$(1+z)^{-1}$ is the cosmological time dilation factor, and
$z_s(S_{\nu, \lim})$ and $z_r(S_{\nu, \lim})$ are the redshifts
corresponding to single and repeating FRBs with a flux $S_{\nu, \lim}$,
respectively.  In the case of repeating FRBs, the formation
probability of a repeating FRB source after a BNS merger ($f_r$) and
the number of repeating bursts during its lifetime ($N_r$) are
multiplied.  Here we assumed that all BNS mergers produce a
non-repeating FRB at the time of merger, and assumed the same beaming
factor for the two populations. These assumptions also affect the
ratio $N_{\rm sFRB}/N_{\rm rFRB}$, and uncertainties about these can
be included in the parameter $f_r$.

No repeating FRBs have been detected by Parkes,
and it implies 
\begin{eqnarray}
  \left.\frac{N_{\rm rFRB}}{N_{\rm sFRB}}\right|_{S_{\nu, \rm lim} = 1 \ {\rm Jy}}
\lesssim 0.1,
\end{eqnarray}
which translates into an upper limit on the product $f_r N_r \lesssim
400$. The parameter $N_r$ can be written as $N_r = \tau_{\rm lt} \, k
\, \zeta$, where $\tau_{\rm lt}$ is the lifetime of a repeating FRB
source, $k$ the repeat rate during the active FRB phase, and $\zeta$
the active duty cycle.  Observationally inferred values are $k\sim 3$
day$^{-1}$ and $\zeta\sim0.3$ \citep{Nicholl2017}, and we get $f_r
\tau_{\rm lt} \lesssim$ 1.2 yr. In order for the lifetime to be
consistent with that discussed in section \ref{sec:repeating-FRBs}, a
weak constraint of $f_r \lesssim 0.1$ is obtained, though there is a
large dependence on model parameters.

In figure \ref{fig: logN-logS}, a $\log{N}$--$\log{S}$ plot for sFRBs
and rFRBs is shown. Both populations show the trend of $N(> S_{\nu,
  \lim}) \propto S_{\nu, \lim}^{-1.5}$ in the bright flux limit, as
expected when cosmological effects are negligible.  The curve of
non-repeating FRBs becomes flat at $S_{\nu, \lim} \lesssim 200$ mJy by
the cosmological effects (cosmic volume and the BNS rate evolution),
but such a behavior is not seen for repeating FRBs because their
redshifts are lower and hence cosmological effects are small.  The
ratio $N_{\rm rFRB}/N_{\rm sFRB}$ is also plotted in the figure \ref{fig: logN-logS}.  This rapidly increases with improving
sensitivity at $S_{\nu, \lim} \lesssim$ 1 Jy, because the cosmological
effects work only on non-repeating FRBs. This gives a possible
explanation for the fact that the only repeating FRB was discovered by
Arecibo that has a better flux sensitivity than Parkes.

\section{Discussion and Conclusions}
\label{sec: Conclusions}
In this paper, we investigated BNS mergers as a possible
origin of both repeating and non-repeating FRBs. 

Non-repeating and bright FRBs mostly detected by Parkes may be
produced at the time of a BNS merger, but the environment around the
merger may be polluted by dynamical ejecta, which would prohibit radio
signals to propagate.  We therefore investigated the BNS merger
environment using a general-relativistic hydrodynamical simulation. It
was found that a significant mass ejection that can be resolved by the
current simulation occurs about 1 ms after the merger, and hence there
is a time window of about 1 ms in which the magneto-rotational energy
production rate has become the maximum to produce an FRB emission and
the environment is not yet polluted.  This also gives a possible
explanation for the observed short duration ($\lesssim$1 ms) of
non-repeating FRBs.

A fraction of BNS mergers may leave a stable remnant neutron star, and
such an object may produce faint and repeating FRBs like FRB 121102
detected by Arecibo, after the environment becomes clear for radio
signals. We showed that the environment becomes clear on a time scale
of order years, and after that FRB activities would become weaker on a
similar time scale by the pulsar spin-down. The persistent radio
emission of FRB 121102 can be explained by a pulsar wind nebula
energized by the remnant neutron star. The expected radio source size
is marginally consistent with the observational upper limit, implying
that a source size evolution may be observed in the future.  DM
expected for the radio emitting nebula is smaller than the
observational estimate of DM from the host galaxy of FRB 121102, and
the nebula RM is not significantly larger than those measured in some
FRBs.  Compared with the supernova scenario for young neutron stars to
produce repeater FRBs, the BNS merger scenario predicts a shorter time
scale for the appearance after the merger (or supernova) and a shorter
active lifetime as a repeating FRB source.  The environment around the
young neutron star is more transparent with smaller DM and RM, while
the source size of persistent radio emission is larger. Especially,
the expected large RM implies that the supernova scenario cannot be
applied to all FRBs because some FRBs show small RM.

We then constructed an FRB rate evolution model including these two
populations. Requiring that the discovery rate of a repeating FRB
source is less than 10\% of that for non-repeating FRBs at the search
sensitivity of Parkes, the lifetime of repeating FRB sources
$\tau_{\rm lt}$ is constrained as $f_r \tau_{\rm lt} \lesssim$ 1.2 yr,
where $f_r$ is the fraction of BNS mergers leaving a remnant neutron
star that is stable on a time scale longer than $\tau_{\rm lt}$.  Then
we obtain $f_r \lesssim 0.1$ from $\tau_{\rm lt} \sim$1--10 yrs
obtained in section \ref{sec:repeating-FRBs}, which is not a strong
constraint because it is an order-of-magnitude estimate.  Since
non-repeating FRBs are brighter and hence more distant at a given
sensitivity, the slope of FRB source counts (log$N$-log$S$) is flatter
than that of repeating FRBs.  Therefore the relative ratio of
repeating to non-repeating FRB source counts should rapidly increase
with improving search flux sensitivity. This gives a possible
explanation to the fact that the only repeating FRB 121102 was
discovered by the most sensitive search using Arecibo, and such a
trend can be confirmed with more FRBs detected in the future.  It
should be noted that this trend is expected even if repeating FRBs
originate from supernovae rather than BNS mergers.

In addition to some predictions already mentioned above, the following
predictions can be made based on our hypothesis.  Originating from BNS
mergers, both repeating and non-repeating FRBs should be found both in
star-forming and elliptical galaxies.  FRB 121102 was found in a dwarf
star forming galaxy with low metallicity, and this may favor the SLSN
scenario. However, a strong conclusion cannot be derived from only one
event; BNS mergers should also occur in such galaxies.  It is
plausible that FRBs showing negligible RM (FRBs 150807 and 150215)
occurred in quiescent galaxies such as elliptical galaxies.

If non-repeating FRBs are produced at the time of BNS mergers, the BNS
merger rate must be close to the high end of the possible range
discussed in the literature, and gravitational wave from a BNS merger
should be detected soon by LIGO/VIRGO/KAGRA. A non-repeating FRB can
in principle be detected coincidentally with gravitational wave from a
BNS merger, but a wide-field FRB search covering a considerable
fraction of all sky will be the key. If the location of a BNS merger
detected by gravitational wave is accurately determined by
electromagnetic wave counterparts, there is a good chance of
discovering repeating FRBs $\sim$1--10 years after the merger, though
the probability of leaving a stable neutron star depends on EOS of
nuclear matter.  A repeating FRB may also be found 1-10 years after a
non-repeating FRB or a short gamma-ray burst (GRB), but they are
generally more distant than BNS mergers detected by gravitational
waves and hence repeating FRBs may be too faint to detect.

Though the fraction of BNS mergers leaving a stable neutron star is
currently highly uncertain, gravitational wave observations may
constrain the nuclear matter EOS in the near future (\cite{Lattimer&Prakash2007}).  Such constraints would
be useful to examine the validity of our scenario for repeating FRBs.
If a repeating FRB is detected after a BNS merger, it would be an
unambiguous proof of a surviving remnant neutron star, which would
give an independent constraint on EOS.  Another possible signature of
a surviving neutron star is a persistent radio emission from the
pulsar wind nebula like FRB 121102, or that from interaction between
ejecta and ISM (\cite{Horesh2016}).

Finally we comment on some intriguing recent observational studies.
\citet{Ofek2017} reported 11 luminous radio sources in nearby ($<$ 108
Mpc) galaxies with offsets from the nucleus, whose luminosities are
similar to the persistent source associated with FRB 121102. The
number density of these is $\sim 5\times10^{-5} \ {\rm
  Mpc^{-3}}$. Using the typical lifetime of repeating FRBs (10 yrs) in
our hypothesis, a birth rate of $\sim 5\times10^{3} \ {\rm yr^{-1}
  \ Gpc^{-3}}$ is inferred, which is interestingly similar to the
non-repeating FRB rate and the high end of the possible BNS merger
rate range. Furthermore, 2 of the 11 sources are in galaxies of old
stellar population (passive and elliptical), which cannot be produced
from young stellar populations.

\citet{Perley2017} reported a new radio source (Cygnus A-2) at a
projected offset of 460 pc from the nucleus of Cygnus A ($z = 0.056$),
which was detected in 2015 but was not present until 1997. The origin
of this source is not yet clear, and a repeating FRB was not discussed
as a possible origin in \citet{Perley2017}.  However we noticed that
the unusually bright radio luminosity as a supernova, $\nu
L_{\nu}\approx6\times10^{39} \ {\rm erg \ s^{-1}}$, is interestingly
similar (within a factor of a few) to that of the persistent radio
emission of FRB 121102, while Cygnus A-2 is about three times closer
to us. The luminosity and the appearance time scale imply that Cygnus
A-2 may also be a pulsar wind nebula produced by a BNS merger remnant,
and a radio monitoring of this may lead to a discovery of another
repeating FRB source.

%%%%%%%%%%%%%%%%%%%%%%%%%%%%%%%%%%%%%%%

%%%%%%%%%%%%%%%%%%%%%%%%%%%%%%%%%%%%%%%

\begin{ack}
We thank the anonymous referee for useful comments that improved the paper. We also thank Kazumi Kashiyama, Koutarou Kyutoku, Masaru Shibata, and Toshikazu Shigeyama for discussion and support to this work. 
SY was supported by the Research Fellowship of the Japan Society for
the Promotion of Science (JSPS) (No. {\rm JP17J04010}).  SY was also
supported by a grant from the Hayakawa Satio Fund awarded by the
Astronomical Society of Japan. TT was supported by JSPS KAKENHI Grant
Numbers JP15K05018 and JP17H06362.  KK was supported by JSPS KAKENHI
Grant Numbers JP16H02183, JP15K05077, and JP17H06361, and by a post-K
computer project (Priority issue No. 9) of MEXT, Japan.  Numerical
computations were performed on K computer at AICS (project number
hp160211 and hp170230), on Cray XC30 at cfca of National Astronomical
Observatory of Japan, FX10 and Oakforest-PACS at Information
Technology Center of the Unversity of Tokyo, and on Cray XC40 at
Yukawa Institute for Theoretical Physics, Kyoto University.
\end{ack}

%%%
% See the manual for the detail.
%%%

\end{document}